\documentclass[3p]{elsarticle}

\usepackage{lineno,hyperref}
\modulolinenumbers[5]

\makeatletter
\def\ps@pprintTitle{%
 \let\@oddhead\@empty
 \let\@evenhead\@empty
 \def\@oddfoot{}%
 \let\@evenfoot\@oddfoot}
\makeatother

\journal{Journal of Magnetism and Magnetic Materials}
\usepackage{color,soul}
\usepackage{graphicx}
\usepackage{bm}
\usepackage{dcolumn}
\usepackage{sidecap}
\usepackage{amsmath}
\usepackage{epstopdf}
\usepackage[version=4]{mhchem}

\newcommand{\SZ}{\langle S_Z \rangle}

\newcommand{\LZ}{\langle L_Z \rangle}

\begin{document}
\title{Intercalated Rare-Earth Metals under Graphene on SiC}

\author[ames]{Nathaniel A. Anderson}
\author[ames]{Myron Hupalo}
\author[aps]{David Keavney}
\author[ames]{Michael Tringides}
\author[ames]{David Vaknin}

\address[ames]{Ames Laboratory, Department of Physics and Astronomy, Iowa State University, Ames, Iowa 50011, United States}
\address[aps]{X-ray Science Division, Advanced Photon Source, Argonne National Laboratory, Lemont, Illinois 60439, United States}



\date{\today}

\begin{abstract}
 Intercalation of rare earth metals ($RE$ = Eu, Dy, and Gd) is achieved by depositing the $RE$ metal on graphene that is grown on silicon-carbide (SiC) and by subsequent annealing at high temperatures to promote intercalation. STM images of the films reveal that the graphene layer is defect free and that each of the intercalated metals has a distinct nucleation pattern.  Intercalated Eu forms nano-clusters that are situated on the vertices of a Moir{\`e} pattern, while Dy and Gd form randomly distributed nano-clusters.  X-ray magnetic circular dichroism (XMCD) measurements of intercalated films  reveal the  magnetic properties of these  $RE$'s nano-clusters. Furthermore, field dependence and temperature dependence of the magnetic moments extracted from the XMCD show paramagnetic-like behaviors with  moments that are generally smaller than those predicted by the Brillouin function.  XMCD measurements of $RE$-oxides compared with those of the intercalated $RE$'s under graphene after exposure to air for months indicate that the graphene membranes protect these intercalants against oxidation.
\end{abstract}

\maketitle

\section*{1. Introduction}
Metal intercalation under graphene on various substrates is by now common practice\cite{Premlal2009,Weser2011,Sandin2012,Schumacher2013,Voloshina2014,Schumacher2014,Ichinokura2016,NarayananNair2016,Sung2017,Zhang2017,Huttmann2017}.  Indeed, it has been demonstrated that rare earth metals ($RE= $ Nd, Sm, Dy, Er and Yb, and Eu, for example) deposited on graphene grown on SiC or Ir substrate can intercalate under the graphene lattice by annealing pre-deposited metals at elevated temperatures\cite{Premlal2009,Schumacher2013,Sung2017,Anderson2017}.  
Adsorbed metals on graphene have also been used to take advantage of special growth or to mutually modify the electronic properties of graphene and the adsorbed metal atoms on it. Single Co atoms  situated in the sixfold hollow site of graphene exhibit magnetic anisotropy considering the weak spin-orbit coupling with the substrate suggesting magnetic impurities on graphene can be used in spintronics applications.\cite{Donati2016} It has also been demonstrated that  high density magnetic islands can be grown on graphene.\cite{Liu2011,Liu2012,Liu2014,Anderson2017}
  In addition to modifying the electronic properties of graphene, intercalated metals can be protected to a certain degree against chemical reaction with the environment\cite{Dedkov2008a,Dedkov2008,Bohm2014,Bunch2008,Anderson2017b}. $RE$ metals have been intercalated into bulk graphite to form graphite intercalation compounds (GICs) for quite sometime, mostly forming an inplane super structure $\sqrt{3}\times\sqrt{3}R30^\circ$ at a saturated stoichiometry  C$_6RE$ (i.e., stage I $RE$--GIC).\cite{Guerard1974,Guerard1975,Hagiwara1996} Such magnetic metal graphene heterostructures have attracted  interest for applications in high frequency electronics, spintronics, and photovoltaic devices\cite{Zhang2014a}.  

 Here, we report on the magnetic properties of intercalated Dy and Gd atoms between graphene and the SiC buffer-layer by employing synchrotron X-ray magnetic circular dichroism (XMCD). The location of the intercalated atoms, whether between graphene and buffer or buffer and SIC, can affect the sample properties and is still an open question. Different phases of the metals have been prepared. The phase studied in this experiment is the phase formed at the lowest preparation temperature where the intercalated atoms initially occupy the top layer by virtue of being kinetically more accessible. We also, report on the chemical stability of the buried Eu layer as samples are exposed to air over a period of days to months.  Intercalation of Eu between Ir substrate and graphene (prepared by CVD) reveals that the structure and magnetic properties of the intercalated Eu depend on the coverage which does not seem to affect the graphene electronic structure\cite{Schumacher2014}.  However, a recent study shows that Eu intercalation between graphene and the SiC buffer layer modifies the $\pi-$band of graphene significantly\cite{Sung2017}.  
 
\section*{2. Methods}
\begin{figure}\centering \includegraphics [width = 3.3 in] {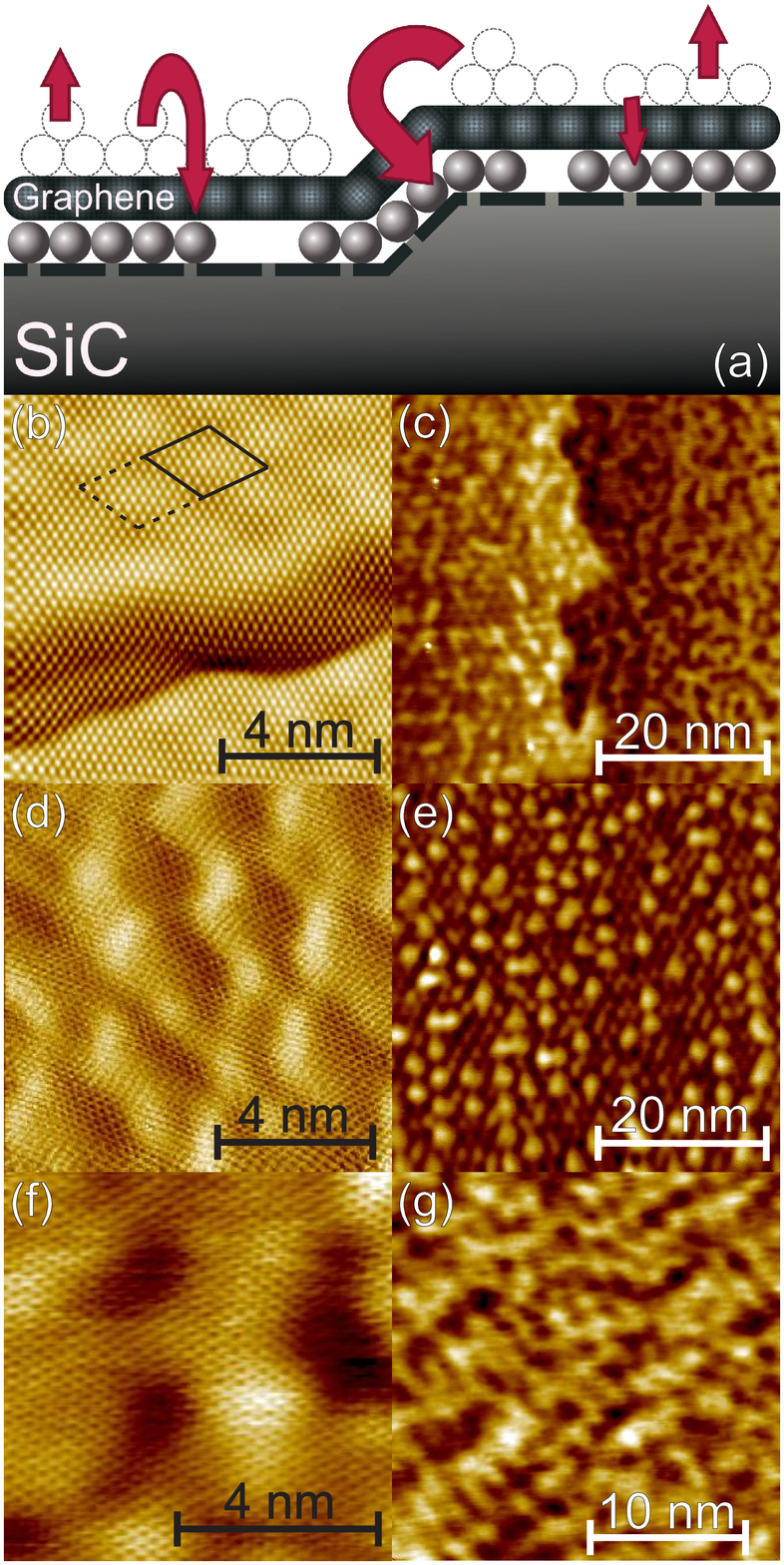}
\caption{(Color Online) (a) Schematic illustration of intercalation of $RE$'s on graphene.  During the annealing process, some atoms penetrate the graphene and intercalate and some just evaporate. (b) STM image of a pristine graphene on a SiC(0001) surface showing the well established 6$\times$6 superlattice. (c)  Dy intercalated under graphene, (d) high resolution image of graphene lattice with Eu intercalated underneath, (e) Eu intercalated under graphene, (f)  high resolution image of graphene lattice with Gd intercalated underneath, and (g) Gd intercalated under graphene.}
\label{fig:STM}
\end{figure}

For the graphene substrate, we use 6H-SiC(0001) wafers (purchased from Cree, Inc) that are graphitized in ultra-high vacuum (UHV, P $\approx1\cdot10^{-10}$ Torr) by direct current heating  to $\sim$1300 C measured with an infrared pyrometer. Figure \ \ref{fig:STM}b shows the pristine graphene layer with the Moir{\`e} superstructure prior to intercalation.
To achieve intercalation, crystalline $RE$ metal (in this case Eu, Dy, or Gd) islands are deposited and grown on the graphene surface by a molecular beam source at a flux rate of 0.1 - 0.2  monolayers (ML)/min\cite{Hupalo2009}. The $RE$ source is degassed during the bake-out for several hours, so during deposition the pressure remains below $1.6\cdot10^{-10}$ Torr. Then, intercalation proceeds by a slow annealing process which differs for each $RE$ element as detailed in the SI. Post-intercalation STM images in Fig. \ \ref{fig:STM}c-g show the intercalated Dy, Eu, and Gd, respectively. Higher resolution STM images (Fig.\ \ref{fig:STM} d and f)  show  that bright spots of the intercalated metal are now situated below the graphene\cite{Anderson2017b}.

XMCD measurements are performed at the  4-ID-C beamline at the Advanced Photon Source (Argonne National Laboratory) in a vacuum chamber of a cryo-magnet  with a high magnetic field ($<6$ T) produced by a split-coil superconducting magnet. Field dependence of the XMCD spectra are collected in helicity-switching mode in external magnetic fields applied parallel to the x-ray wave vector at energies that cover the $M_4$  (Dy: 1333 eV; Gd: 1221 eV) and $M_5$  (Dy: 1292 eV; Gd: 1189 eV) binding energies.   The X-ray incident angle is fixed at $\sim20\pm2$ degrees with respect to the sample surface and x-ray absorption spectroscopy (XAS) signals are collected by total electron yield (TEY). For data analysis and normalization, the individual XAS, $\mu_+$ and $\mu_-$, are normalized by their respective monitors to compensate for incident-beam intensity variations. For the initial background subtraction, the XAS ($\mu_+$ and $\mu_-$) has a flat value subtracted such that the lowest energy (i.e. sufficiently far from the edge) is at 0 intensity, removing both background and offsets due to the beam. The total XAS ($\mu_++\mu_-$) is then scaled by a factor such that its maximum intensity is 1. That scale factor is then used to also scale the individual ($\mu_+$ and $\mu_-$) XAS. The XMCD signal is obtained from the difference between two XAS spectra of the left- and right-handed helicities, $\mu_+$  and $\mu_-$. More details on data reduction is provided elsewhere\cite{Anderson2017,Anderson2017b}.

\section*{3. Results and Discussion}

\subsection*{3.1 Europium}
We have recently reported our results on intercalated Eu under graphene\cite{Anderson2017b}. Briefly, the intercalated Eu forms nano-clusters at nucleation sites  located at the vertices of the 6$\times$6 superstructure formed by the graphene on SiC (Fig.\ \ref{fig:STM}e). Fig. \ \ref{fig:STM}d is a high resolution image in differential mode to suppress the corrugation from the Eu and enhance the graphene lattice on top. Using  XMCD measurements at the Eu M$_{4,5}$ edges show that the intercalant is Eu$^{2+}$ with paramagnetic-like behavior, as determined from the magnetic field and temperature dependencies of the XMCD.  Also, no evidence of ferromagnetism due to EuO or antiferromagnetism due to \ce{Eu2O3} has been detected in the XMCD as a function of temperature indicating that the graphene layer protects the intercalated metallic Eu against oxidation over months of exposure to atmospheric environment\cite{Anderson2017b}.

\subsection*{3.2 Dysprosium}

\begin{figure}\centering \includegraphics  [width = 76 mm] {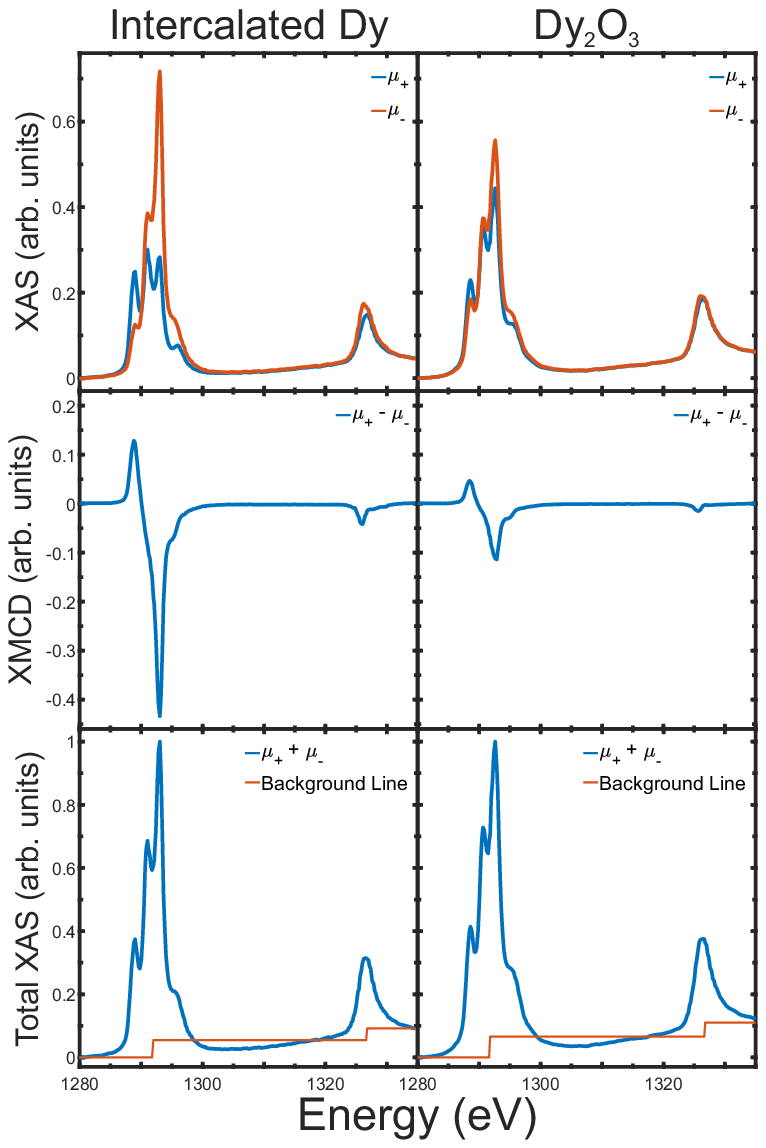}
\caption{(Color Online) The $\mu_+$ and $\mu_-$ XAS, XMCD, and total XAS of the Intercalated Dy (left) and polycrystalline \ce{Dy2O3} (right) at $H =5$T and $T=15$ K.}
\label{fig:DyXAS}
\end{figure}

Unlike the Eu clusters, Dy displays a very different intercalation pattern. The intercalated Dy forms amorphous clusters as shown in Fig.\ \ref{fig:STM}c. The clusters have no apparent pattern or uniformity. Figure\ \ref{fig:DyXAS} shows the XAS at the Dy $M_4$ and $M_5$ edges for the two circularly polarized X-ray beams at $T =15$ K and $H = 5$ T for  intercalated Dy (left panel) and for polycrystalline \ce{Dy2O3} (right panel). 

\begin{figure}\centering \includegraphics [width = 76 mm] {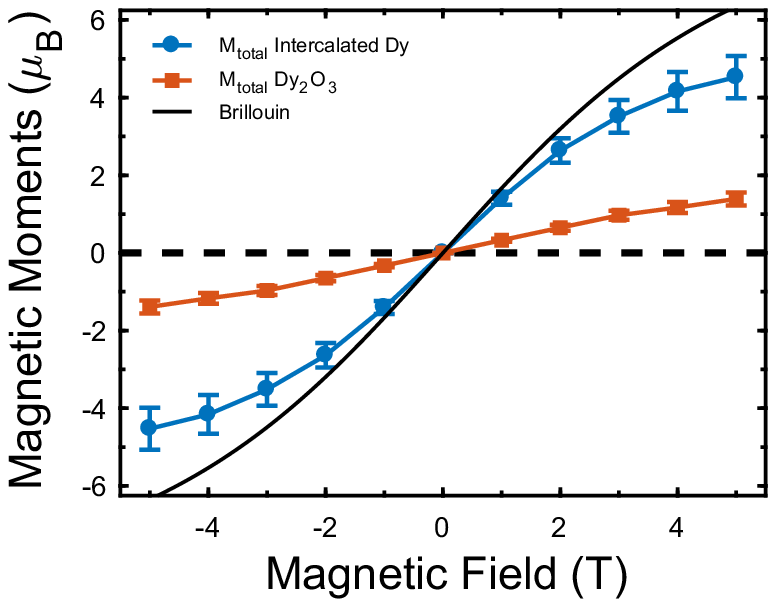}
\caption{(Color Online) The magnetic field dependence of the Intercalated Dy (blue circle) and \ce{Dy2O3} (red square) at T=15K. The Brillouin Function for Dy$^{3+}$ (black line) is also included for comparison.}
\label{fig:DyField}
\end{figure}

Both samples exhibit similar total XAS signals characteristic of Dy$^{3+}$, however, with distinct differences in the shapes of the $\mu_+$ and $\mu_-$ signals that are further emphasized in the XMCD pattern ($\mu_+-\mu_+$). In particular, the intercalated sample has a significantly stronger signal than that of the \ce{Dy2O3} providing qualitative evidence that the intercalated Dy has a larger magnetic moment. We note that metal Dy is paramagnetic (PM) at room temperature and undergoes a first order magnetic transition from paramagnetic to incommensurate helical structure with a temperature dependent pitch at $T_{\rm H}=179$ K and a second transition at $T_{\rm C} \approx 88$ K from the helical to a ferromagnetic (FM) structure\cite{Wilkinson1961,Yu2015}.  Using sum rules appropriate for Dy$^{3+}$ (similar to those used for the XMCD from Dy islands on graphene\cite{Anderson2017}),  we extract the total magnetic moment as a function of magnetic field as shown in Fig.\ \ref{fig:DyField}.  Also shown in Fig.\  \ref{fig:DyField} is the field dependence of the Dy moment in \ce{Dy2O3} where it is evident that  the magnetic moment of the intercalated Dy is more than 3 times  that of the \ce{Dy2O3} however, slightly smaller than that expected for the theoretical paramagnetic value  predicted by the Brillouin function for Dy$^{3+}$. The field dependence of the XMCD does not show hysteresis or anisotropy that is characteristic of ferromagnetic systems. We note that  Eu$^{3+}$ in \ce{Eu2O3} has no magnetic moment  based on Hund's rules (i.e., $J=0$) , and thus has no XMCD signal\cite{Anderson2017b}, while \ce{Dy2O3} and \ce{Gd2O3} have a non-zero $J$ exhibiting detectable XMCD signal.

\begin{figure}\centering \includegraphics [width = 76 mm] {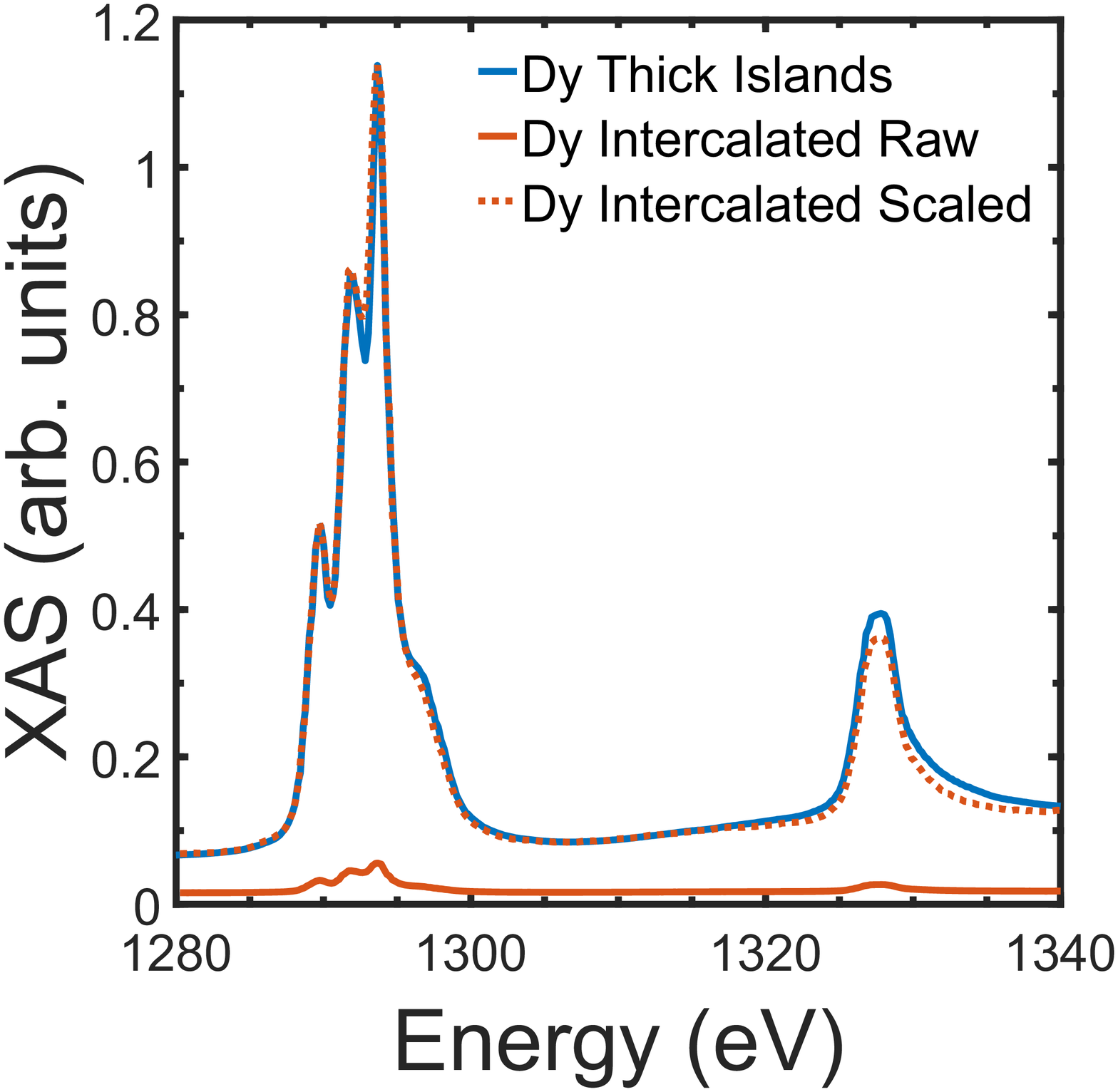}
\caption{(Color Online) The XAS of a Thick Dy Islands sample (blue, 27ML) compared to the XAS of the Intercalated Dy (solid red). The Intercalated Dy Scaled (dashed red) is related to the raw spectra via: $Scaled=26.6\cdot Raw-0.35$.}
\label{fig:DyScale} 
\end{figure}

A common challenge with intercalation is the determination of the coverage that is achieved post-intercalation. After deposition and before intercalation, it is straight forward to measure the coverage to obtain an upper limit for the amount of material, however, the annealing required for intercalation causes evaporation of some material. When the $RE$ element is between the graphene and buffer layer, it is no longer easy to distinguish if clusters are a single layer or multiple layers, thus making determination of the coverage possible, but unreliable. To estimate the intercalant content we compare the XAS of the intercalated Dy to the XAS of a non-intercalated sample of Dy islands of a known coverage (27 ML; coverage is determined by the flux rate  and confirmed by integrating the volume of different islands (multiply island height times area and use the density of fcc Dy to get the amount\cite{Anderson2017}. The XAS of the intercalated and islands are plotted on the same graph and scaled and shifted such that it overlays  the Dy island signal (Fig.\ \ref{fig:DyScale}), using the following scaling $XAS_{Island} \approx 26.6XAS_{Int}-0.35$ where  $XAS_{Islands}$ and $XAS_{Int}$ are the signal from the Dy nano-islands and the intercalated. With this, we estimate the average nominal coverage of the intercalated Eu at $\frac{27}{26.6} \approx 1.0$ ML, which is consistent with the nominal amounts of monolayers deposited.

\subsection*{3.3 Gadolinium}

\begin{figure}\centering \includegraphics  [width = 76 mm] {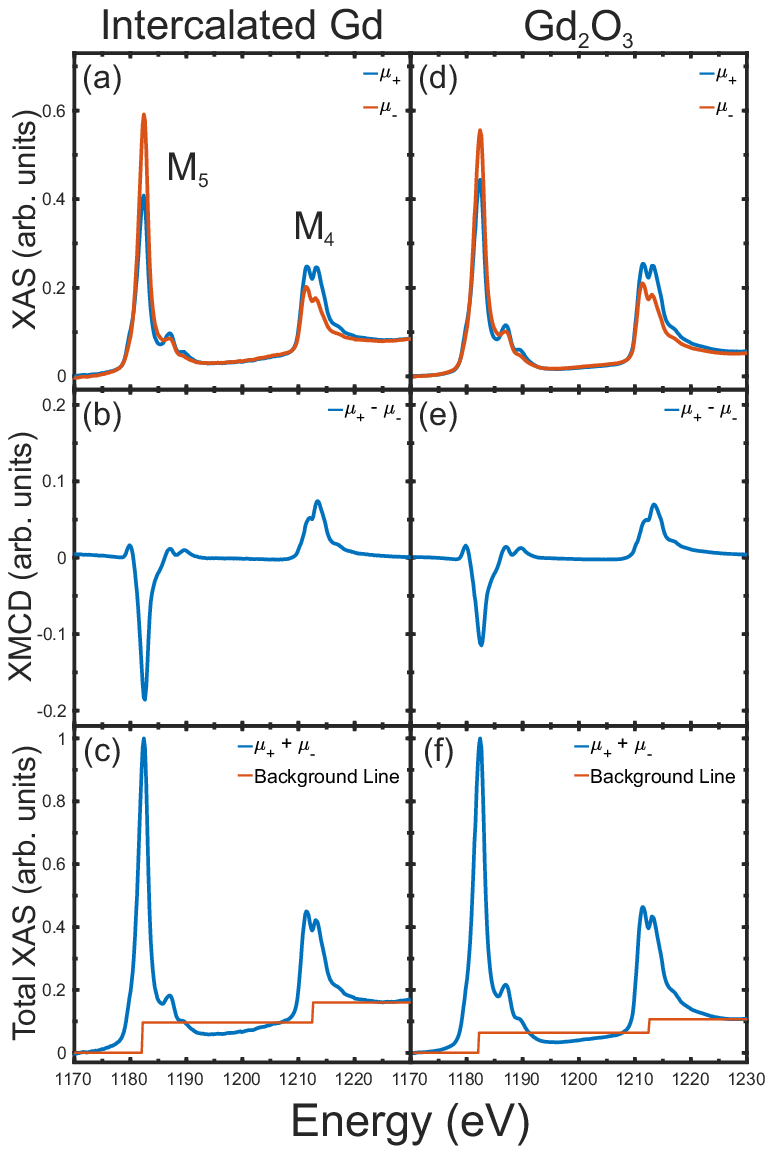}
\caption{(Color Online) The $\mu_+$ and $\mu_-$ XAS, XMCD, and total XAS of the Intercalated Gd (left) and polycrystalline \ce{Gd2O3} (right) at $H=5$ T and $T=15$ K. The spectra are nearly identical, with the only difference being the intensity of the XMCD.}
\label{fig:GdXAS}
\end{figure}

\begin{figure}\centering \includegraphics [width = 76 mm] {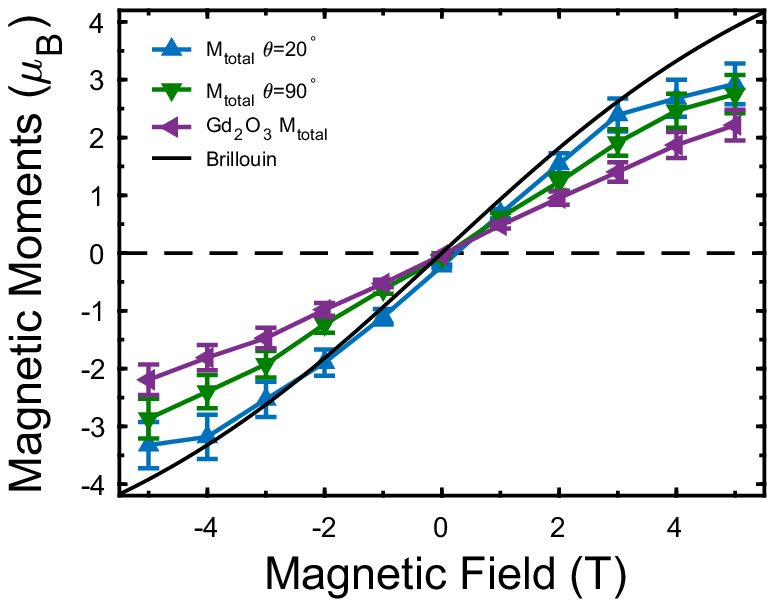}
\caption{(Color Online) The magnetic field dependence of the Intercalated Gd compared with \ce{Gd2O3}. The up and down triangles show the $M_{total}$ for Intercalated Gd with the beam at an angle of 20$^\circ$ and 90$^\circ$ from the surface. The Brillouin Function for Gd$^{3+}$ (black line) is also included for comparison.  }
\label{fig:GdField}
\end{figure}

Just like the intercalation pattern of Dy, the Gd forms random nano-clusters with no ordered pattern (Fig.\ \ref{fig:STM}f-g). Figure\ \ref{fig:GdXAS} shows the XAS at the Gd $M_4$ and $M_5$ edges for the two circularly polarized X-ray beams at $T =15$ K and $H = 5$ T for the intercalated Gd (left) and for polycrystalline \ce{Gd2O3} (right). 
Whereas the total XAS of the two samples are nearly identical,  the intercalated Gd has a stronger XMCD signal indicating a  larger magnetic moment compared to the oxide as shown in the field dependent XMCD (Fig.\ \ref{fig:GdField}). The similar XAS patterns show that the Gd is in the common Gd$^{3+}$ state in the metal, as expected. Since Gd$^{3+}$ has the same electron configuration as Eu$^{2+}$,  their XMCD sum rules are the same resulting in an expected orbital moment, $\LZ = 0$, , and a total moment, $M_{total}$, which is equal to twice the spin moment, $\SZ$. \cite{Anderson2017a}. The field dependence shown in Fig.\ \ref{fig:GdField} does show a strong total moment which appears paramagnetic but does not saturate. The measured $\LZ$ moment appears to fluctuate around 0 such that it is $0\pm0.5 \mu_B$ (See SI).
In addition to  measuring the XMCD at incidence angle of 20$^\circ$, the Gd sample was also measured at 90$^\circ$ (X-rays normal to the sample) to check for possible anisotropy. As seen in Fig.\ \ref{fig:GdField}, there is a small insignificant difference in the $M_{total}$ from the 20$^\circ$ and 90$^\circ$ at low fields. This is also the same observation for a second high concentration Gd sample (See SI). Both angles have slightly higher moments than that of the \ce{Gd2O3}, but seem to display a moment that is slightly lower than  predicted by the Brillouin function (solid black line).

\begin{figure}\centering \includegraphics [width = 76 mm] {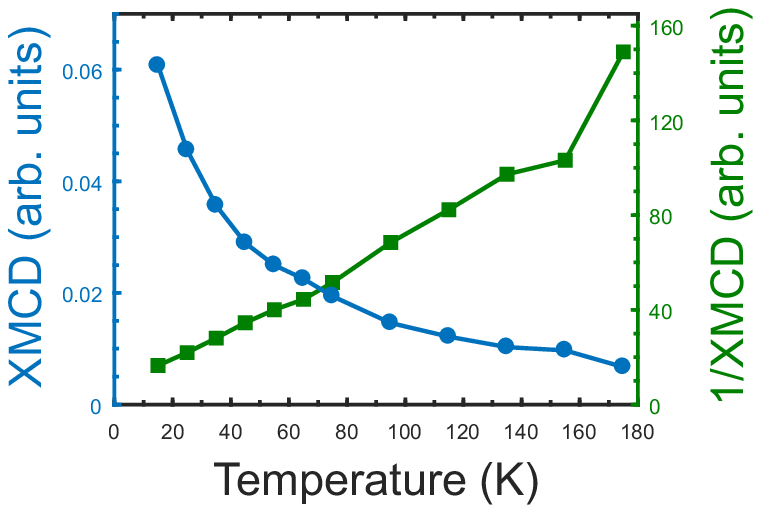}
\caption{(Color Online) The temperature dependence of Intercalated Gd. Each point is the peak intensity of the XMCD on the $M_5$ edge. The 1/XMCD dependence is  linear with an anomaly at the final temperature of 175 K.}
\label{fig:GdTempPeak} 
\end{figure}

For the intercalated Gd, a temperature dependence of the XMCD was also conducted. Figure \ \ref{fig:GdTempPeak} shows the temperature dependence of the XMCD from 15K to 175K determined by integrating over the peak of the the $M_5$ XMCD. The inverse of the peak (1/XMCD) is also plotted which is roughly equivalent to 1/$\chi$. With a linear temperature dependence whereby $\theta$ of the Curie-Weiss law is slightly below 0, indicating anti-ferromagnetic intercations. There is an anomaly at 175K where the 1/XMCD has a spike, but without more data at higher temperatures it is unclear whether this is real or an outlier. Both the field and temperature dependence of the intercalated Gd indicate that it is paramagnetic at low temperatures. This differs from the properties of bulk Gd which is FM with a transition temperature $T_{\rm C} \approx 290$ K\cite{Elliott1953,Will1964}, but is in line with the other intercalated $RE$'s that do not show collective phenomena, i.e. ferro or antiferromagnetism.  Since the structures of bulk Gd is likely different than that of the intercalated clusters, the difference in magnetic behavior is expected. Even if the intercalated layer is ordered, the  local environments in bulk which includes adjacent layers is expected to exhibit different magnetic behavior than the intercalated  Gd layer.


\section*{4. Conclusions}
Using XMCD measurements of graphene intercalated $RE$'s on silicon carbide reveal similar surface paramagnetic properties in these single layer $RE$'s. STM images (Fig.\ \ref{fig:STM}c-g) show that some of the metals have a distinct nucleation pattern after intercalation, whereby Eu forms nano-clusters that are located on a  super lattice creating a Moir{\`e} pattern with the graphene, and  the Dy and Gd clusters are randomly distributed.  Furthermore, the field dependence and temperature dependence of the $RE$ magnetic moments extracted from the XMCD show paramagnetic behavior, while having slightly lower moments than predicted by the corresponding Brillouin function. Our results also show that the intercalants exhibit magnetic moments that are significantly different than those of $RE$s' oxides suggesting that the intercalated species are metallic and that the graphene membrane protects the intercalated films against oxidation over prolonged periods of time.

\section*{Acknowledgments}
Ames Laboratory is operated by Iowa State University by support from the U.S. Department of Energy, Office of Basic Energy Sciences, under Contract No. DE-AC02-07CH11358. Use of the Advanced Photon Source, an Office of Science User Facility operated for the U.S. Department of Energy (DOE) Office of Science by Argonne National Laboratory, is supported by the U.S. DOE under Contract No. DE-AC02-06CH11357.

\section*{References}
\bibliography{IntercalatedDyGd.bbl}
\end{document}